\documentclass[twocolumn,twoside]{article}
\usepackage{hyperref}

\begin{document}

\thispagestyle{plain}
\markboth{\rm ACTIVE GALACTIC NUCLEI AND TRANSFORMATION OF
DARK MATTER ...}{\rm GRIB, PAVLOV}

\twocolumn[
\begin{center}
{\LARGE \bf Active Galactic Nuclei and Transformation

\vspace{10pt}
of Dark Matter into Visible Matter${}^1$}
\vspace{12pt}

{\Large {A. A. Grib}${\,}^{2,\, 3 {}^{*}}$ \ and \
{Yu. V. Pavlov}${\,}^{2,\,4 {}^{**}}$} \vspace{12pt}

{\it ${}^2$A. Friedmann Laboratory for Theoretical Physics,\\
Griboedov kanal 30/32, St.\,Petersburg 191023, Russia}
\vspace{4pt}

{\it ${}^3$Russian State Pedagogical University (The Herzen University),\\
Moyka-river Emb. 48, St. Petersburg 191186, Russia}
\vspace{4pt}

{\it ${}^4$Institute of Mechanical Engineering, Russian Acad. Sci.,\\
Bolshoy pr. 61, St. Petersburg 199178, Russia}
\end{center}
\vspace{17pt}
 {\bf Abstract.}
The hypothesis that dark matter is converted into visible particles
in active galactic nuclei is investigated.
If dark matter consists of stable superheavy neutral particles and
active galactic nuclei are rotating black holes, then, due to the Penrose
process, superheavy particles can decay into unstable particles with
larger mass, whose decay into quarks and leptons leads to events in
cosmic rays observed by the Auger group.
Similar processes of decay of superheavy particles of dark matter into
visible matter occurred in the early Universe.
Numerical estimates of the processes in active galactic nuclei and in
the early Universe are given.

\vspace{11pt}
\vspace{27pt}
]

{\centering  \section{\uppercase{Introduction}}}

\footnotetext[1]{E-mail: \, andrei\_grib@mail.ru \\
\hspace*{7pt}${}^{**}$ \hspace{-6pt}  E-mail: \, yuri.pavlov@mail.ru \\
\hspace*{10pt}${}^{1}$ \hspace{-5pt}
Plenary talk given at the International Conference\\
\hspace*{12pt} RUSGRAV-13, June 23-28, 2008, PFUR, Moscow}

    Recently the Auger group for observation of cosmic rays
claimed~\cite{Auger07} that 27 events with ultra high energy
$E \ge 57 \times 10^{18}$\,eV were observed.
Some correlation of these ultra high energy cosmic rays (UHECR) with the
distribution of nearest active galactic nuclei (AGN) was also observed.

    In our previous papers~\cite{GrPvAGN}
we tried to explain these events by the hypothesis investigated by
us~\cite{GrPv} and some other authors~\cite{DubKhlop}
that dark matter consists mainly of neutral stable superheavy particles
with masses of the order of the Grand Unification (GU) scale
$M_U=10^{14} - 10^{15} $\,GeV.
    The argument for this was due to numerical estimate obtained
earlier~\cite{GMM,GD}
that gravitation of the early expanding Universe could produce the observable
number of visible particles (the Eddington number) if it first created from
vacuum superheavy particles which then decayed into visible ones
but part of them survived till today as dark matter.

    Supposing that similar to $K^0$-meson theory, superheavy particles can
exist in two forms, the long-living component existing today and the
short-living one, which decayed in the early Universe,
it is interesting to investigate the idea that in AGN the process of
converting long-living components into short-living ones, decaying into
visible particles, can occur.
    As it was said in~\cite{GrPvAGN},
this can be the Penrose process in rotating black holes~\cite{Penrose69}.
    As it is known, in this process the superheavy particle in ergosphere
of the rotating black hole can decay into two particles:
one with the negative energy, the other with positive energy.
   So it is possible that the particle with positive energy can have
a larger mass than the decaying one.
    Calling these particles $X_l$ and $X_s$, one can have
$m_{X_l} < m_{X_s}$.
In this sense, such situation is different from the usual $K^0$-meson theory
with $m_{K_l} > m_{K_s}$.
    Then $X_s$ decays into quarks and leptons of superhigh energies.
The flow of these particles escapes the AGN, and some of them come
to the Earth.

    It is clear that in ordinary space, as well as in galactic nuclei
which are not rotating black holes, this process cannot be observed and
dark matter is stable.

    The structure of the paper is the following.
First we analyze the Penrose process in AGN as rotating black holes.
Then we repeat our arguments on the role of superheavy particles in
the early Universe.

\vspace{10pt}
{\centering \section{Superheavy particles as sources of UHECR from active
galaxy nuclei}
\label{UAGN}
}

    Let us give some numerical estimates of the conversion of superheavy
particles into UHECR in active galactic nuclei.
The Auger group detected 27 UHECR with energies higher than
$57 \times 10^{18}$\,eV.
The integrated exposure of the Auger observatory for these data is
$9.0 \times 10^3$\,km${}^2\,$sr\,year.
The Auger group found the correlation of UHECR with nearby active
extragalactic objects~\cite{Auger07}.
There are 318 AGN in the field of view of the
Observatory at distances smaller than 75\,Mpc.
It is easy to see that if these AGN are distributed uniformly and have
the same intensity of UHECR radiation, each of the AGN must radiate
approximately $j=10^{39}$ UHECR per year.
    The distance of propagation of UHECR is limited by
the Greizen-Zatsepin-Kuzmin limit~\cite{GZK} and for a proton with
an energy of the order $8 \times 10^{19}$\,eV this distance cannot be larger
than 90\,Mpc.

    Due to the Auger results, the source of 2 particles of superhigh energy
is the AGN Centaurus~A located at 11 million of light years from the Earth.
It is easy to calculate that for the integral exposure of the Auger observatory
this AGN must radiate $j \approx~3 \times 10^{37}$ UHECR per year.

    There is no well established theory explaining the origin of
cosmic particles with energies higher than $10^{18}$\,eV in AGN.
Our hypothesis is that these UHECR in AGN arise due to superheavy dark
matter particles converted into quarks and leptons at high energies obtained
by them close to the supermassive black hole horizon of the AGN.
Superheavy dark matter particles with a mass
$M=10^{14}$\,GeV $\approx 2 \times 10^{-10}$\,g, fall onto the
black hole, so that if 100 \% of these particles are converted into UHECR,
then it's mass must have the order of $M j \sim 10^{28}$\,g.
    Even if only $\eta = 10^{-4}$ of the total mass of superheavy particles
close to the horizon is converted into ordinary particles, the whole mass
of dark matter $ \Delta m_a = M j/\eta $ needed to obtain the same current
is much lower than the mass of the ordinary matter accreted onto
the black hole, leading to its observed light radiation.

    The source of the UHECR energy is the decay of superheavy particles
at the GU energies into quarks and leptons which, according to our reasoning
in Part~\hyperref[SHP]{3} of our paper, led to the origination of
the baryon charge of the Universe.
    The mass of a superheavy particle is converted into the energy of light
particles whose flow can come from the black hole to the Earth
similarly to photons.
    The black hole plays the role of a cosmic accelerator or
supercollider creating the conditions for transforming the long-living
component of $X$-particles into short living and its decay.

    Now let us evaluate the density of dark matter needed to form
the observed UHECR flow.
    Suppose that dark matter is uniformly distributed with a density typical
for the ordinary matter in central parts of the galaxies,
$\rho = 10^{-20}$\,g/cm${}^3$.
Let us take the typical velocities of dark matter particles at large
distance from the central black hole as $v_\infty \approx 10^8$\,cm/s
(these are stellar velocities in the central parts of galaxies).
The capture cross-section for nonrelativistic particles by
a Schwarzschild black hole
is given by (see Eq.\,(3.9.1) in~\cite{ZelNovTTES})
     \begin{equation}
\sigma_{c} = 4 \pi \left( \frac{c}{v_\infty} \right)^2 r_g^2 \,.
\label{sz}
\end{equation}
Here $r_g$ is the horizon radius of the black hole.

The capture cross-section for a rotating black hole is
of the same order~\cite{ZelNovTTES}.
Taking $M_{BH} = 10^8 M_\odot $ ($M_\odot $ is solar mass),
one obtains for the velocity of the dark matter accretion onto the black hole
$ \Delta m_a = \sigma_c v_\infty \rho \approx 3 \times 10^{28}$\,g/year,
which is consistent with our evaluation of the Auger observation
$j \approx 10^{38}$ UHECR/year.
    This nontrivial coincidence can be an argument for our reasoning.

    One must mention that conversion of dark matter into UHECR is effective
only for objects with large quantity of the diffuse dark matter close to
the black hole.
This situation can occur only in AGN and is improbable for ordinary galaxies.
From~(\ref{sz}) one can see that capture of dark matter by a black hole
is proportional to the square of black hole mass, so that the UHECR flow
from a black hole of stellar masses is negligible.

    We have no observation data for the distribution of dark matter
in central regions of galaxies with AGN.
If one takes for the distribution density of dark matter the numerical
profiles
     \begin{equation}
\rho(r) = \frac{\rho_0}{(r/r_0)^\beta (1+r/r_0)^{3-\beta}}
\label{rNFW}
\end{equation}
with $\beta=1$ for Navarro-Frenk-White profile~\cite{NFW96},
$\beta=1.5$ for Moore profile, $r_0=45$\,kpc,
$\rho_0 =10^{-24}$\,g/cm${}^3$ ~\cite{ABerK},
then one again obtains
$ \Delta m_a \sim 2 \times 10^{28} - 10^{30}$\,g/year.

So our estimate leads to a reasonable number for the accretion of
supermassive dark matter particles onto a black hole.
This dark matter can be considered to be a source of UHECR arising
from the decay of supermassive particles into visible matter close
to the horizon of a supermassive black hole.

    The maximum energy of the observable cosmic rays is not higher
than $3 \times 10^{20}$\,eV.
Note that the absence of observations of cosmic rays with higher energy
does not contradict the hypothesis that the source of UHECR is the decay of
superheavy particles of the mass of the order $M=10^{14}$\,GeV near
the horizon of the central black holes in AGN.
The observable luminosity of AGN is $10^{43} - 10^{48}$\,erg/s
close to the centre.
In~such a dense flow of radiation protons of ultrahigh energy will have
many photo-nuclear collisions leading to large energy decrease
(see, e.g.,~\cite{AstroFizKosLuch}, ch.\,4, \S \,5).

    Now let us discuss the possible physical mechanism of conversion of
dark matter into visible matter in AGN.
It is reasonable to think that AGN, unlike other black holes,
are rapidly rotating supermassive black holes with an angular momentum
close to the critical value.
Then one has the well-known Penrose mechanism~\cite{Penrose69}.
    An incoming particle in the ergosphere decays into two or more particles,
one with negative energy goes inside the black hole while other
particles with the opposite momentum and an energy larger than the incoming
one go to the ambient space.

    A new feature of this process is that a long-living particle of dark matter
can decay into a particle of larger mass which can be identified with
the short-living component $X_s$ decaying into particles of visible
matter --- quarks and leptons, as it had happened according to our scenario
in the early Universe.
If the energy transfer in such process is larger than the GU scale, then
neither the baryon charge nor the conserved charge of dark matter
particles are conserved.
    Note that similar processes of decay into a particle of larger mass
than that of the initial one (forbidden in Minkowski space) could occur in
the early expanding Universe~\cite{Ford82}, \cite{GribKryukov88}.
$CPT$-invariance is broken in such processes as the consequence of breaking
of $T$-invariance in a nonstatic metric.

Let us give some necessary formulas and estimates.
    The Kerr metric of a rotating black hole in the Boyer-Lindquist coordinates
has the form
    \begin{eqnarray}
d s^2 \! = d t^2 \! -\! (r^2 \!\! + a^2 \! \cos^2 \! \theta ) \! \biggl(
\frac{d r^2}{r^2 \! - 2 m r \! + a^2} + d \theta^2 \! \biggr)
\nonumber \\
-\, (r^2 + a^2) \sin^2 \! \theta\, d \varphi^2 -
\frac{2 m r \, ( d t - a \sin^2 \! \theta\, d \varphi )^2}{r^2 + a^2 \cos^2
\! \theta },    \hspace{1pt}
\label{Kerr}
\end{eqnarray}
    where $m$ is the black hole mass, $am$ is the angular momentum
($a < m $).
Here we use units $G=c=1$.
At the event horizon
     \begin{equation}
r = r_H \equiv m + \sqrt{m^2 - a^2} \,.
\label{Hor}
\end{equation}
The surface called the stationary limit surface is given by
     \begin{equation}
r = r_0 \equiv m + \sqrt{m^2 - a^2 \cos^2 \! \theta} \,.
\label{PrSt}
\end{equation}
    The ergosphere is the region of space-time between the horizon and
the stationary limit surface.
Inside the ergosphere, the Killing vector of time translation $(1,0,0,0)$
becomes the spacelike, and the energy of a particle in this region measured
by the observer in infinity can be negative.
    This leads to a possibility of the physical process of getting the energy
from a rotating black hole.
    However there are some constraints on this energy.
    In case of particle decay, one has the Wald inequality
(see, e.g.,~\cite{Chandrasekhar}, \S\,65).

    If $E_u$ is the specific energy (energy/rest-mass) of a particle falling
onto a black hole,
$\varepsilon$ and $v$ are the specific energy and relative velocity of one of
the fragments into which the particle decayed, then
     \begin{equation}
\gamma \left( E_u - |v| \sqrt{E_u^2 -
\left( 1 - \frac{2 m r}{r^2 + a^2 \cos^2 \theta } \right)} \right)
\le \varepsilon ,
\label{nWald}
\end{equation}
    where $ \gamma = 1/ \sqrt{\mathstrut 1 - |v|^2 } $ \ and \
$ r_H < r < r_0 $.

    Let us consider the initial particle falling onto the black hole to be
nonrelativistic far from from the black hole, i.e. $E \approx 1$.
    If the initial particle of mass $M$ decays into two particles one of
which is a light particle with mass $\mu $, then one can take
$ |v| \approx 1 $ and $ \gamma \approx M/ 2\mu $.
So one gets a constraint on the additional energy (mass) of
the other fragment:
    \begin{equation}
\Delta E \le \frac{M}{2} \left( \sqrt \frac{2 m r }{
r^2 + a^2 \cos^2 \theta } \, - 1 \right).
\label{Edop}
\end{equation}
    So one can obtain the following estimate of the additional energy
(mass) obtained in the process of decay of the massive particle into light
and heavier particles in the ergosphere of a black hole with the specific
angular momentum $a$:
    \begin{equation}
\Delta E \le \frac{M}{2} \biggl[ \! \biggl( 1- \frac{1}{2}
\biggl( 1- \sqrt{1- \frac{a^2}{m^2} } \, \biggr)\! \biggr)^{\!\! -1/2}
\!\! -1 \biggr].
\label{EdopMax}
\end{equation}
    The maximal value $ \Delta E_M \approx M (\sqrt{2} - 1 )/2 $ is achieves
for a black hole with an angular momentum close to the critical value $a=m$.
    Full description of the process needs some $S$-matrix theory.

If the mass of short-living $X_s$-particles is larger than the mass of
long living ones, then their creation in the Penrose process and consequently
the appearance of the UHECR in decays of the short living component of
$X$-particles will be obtained only for rapidly rotating black holes.

    So the necessary condition for the Penrose mechanism --- ultra
relativistic movement of decay products as well as condition for
converting of superheavy dark matter particles into quarks and leptons,
i.e., large value of the relative energy-momentum in interaction of these
particles, can be fulfilled in process under consideration.
    Outgoing particles with energy larger than the Grand Unification
scale can collide with other superheavy particles and ordinary matter.
As a result, macroscopic amount of dark matter can be ``burnt'' close
to AGN and create UHECR.
So AGN can work as a large cosmic collider.

\vspace{10pt}
{\centering \section{Superheavy particles in the early Universe}
\label{SHP}
}

     The total number of massive particles created in the
Friedmann radiation-dominated Universe
(with the scale factor $a(t)=a_0\, t^{1/2}$)  inside the horizon is,
as it is known~\cite{GMM},
    \begin{equation}
N=n^{(s)}(t)\,a^3(t)=b^{(s)}\,M^{3/2}\,a_0^3 \ ,
\label{NbM}
\end{equation}
    where $b^{(0)} \approx 5.3 \times 10^{-4}$ for scalar
and  $b^{(1/2)} \approx 3.9 \times 10^{-3}$ for spinor particles.
It occurs that $ N \sim 10^{80} $ for $ M \sim 10^{14} $\,GeV \cite{GD}.
    The radiation dominance at the end of inflation era for dark matter is
important for our calculations.
    If it is dust-like, the results will be different (see further).
    However, this radiation is formed not by our visible particles.
It is quintessence or some mirror light particles, not interacting with
ordinary particles.

    For the time ${t \gg M^{-1}} $, there is a transition era from a
radiation-dominated model to a dust model of superheavy particles,
    \begin{equation}
t_X\approx \left(\frac{3}{64 \pi \, b^{(s)}}\right)^2
\left(\frac{M_{Pl}}{M}\right)^4 \frac{1}{M}  \,,
\end{equation}
    where $M_{Pl} \approx 1,2 \times 10^{19}$\,GeV is the Planck mass.
    If $M \sim 10^{14} $\,GeV,
$\ t_X \sim 10^{-15} $\,s for scalar and
$\ t_X \sim 10^{-17} $\,s for spinor particles.

    Let us call $t_X$ \ the ``early recombination era''.

The formula for created particles in the volume $a^3(t)$
can be written as
     \begin{equation}\label{Na}
N(t)= \left( \frac{a(t_C)}{t_C} \right)^{\! 3} b^{(s)},
\end{equation}
where $t_C=1/M$ is the Compton time, $b^{(s)}$ depends on the form of $a(t)$.
From~(\ref{Na}) one can see the effect of connection of the number of
created particles with the number of causally disconnected parts on the
Friedmann Universe at the Compton time of its evolution.

    For scale factor $a(t)=a_0\, t^{\alpha}$ from Eq.~(\ref{Na})
it follows that $\ N = b^{(s)}\,M^{\, 3(1-\alpha)}\,a_0^3 $.
   Therefore for a dust-like end of inflation era one has
$ N \sim M $, and  the ratio of
the $X$-particles energy density $\varepsilon_X$ to the critical
density $\varepsilon_{crit}$  is time-independent
($\varepsilon_X < \varepsilon_{crit}$ for $M < M_{Pl}$).

   Let us define $d $, the permitted part of long-living
$X$-particles, from the condition: at recombination time $t_{rec} $
in the observable Universe one has
$
d\,\varepsilon_X(t_{rec}) =\varepsilon_{crit}(t_{rec})  \,.
$
    It leads to
\begin{equation}
d=\frac{3}{64 \pi \, b^{(s)}}\left(\frac{M_{Pl}}{M}\right)^2
\frac{1}{\sqrt{M\,t_{rec}}}\, .
\label{d}
\end{equation}
     For $M=10^{13} - 10^{14} $\,GeV one has
$d \approx 10^{-12} - 10^{-14} $ for scalar and
$d \approx 10^{-13} - 10^{-15} $ for spinor particles.
    So the lifetime of the main part or all $X$-particles must be smaller
or equal than $t_X$.

     Earlier we constructed, by analogy with the $K^0$-meson theory,
a model~\cite{GrPv} which can give: \
{(a)} short-living $X$-particles decay at the time
   $\tau_s < t_X $ (more wishful is
   $\tau_s \sim t_C \approx 10^{-38} - 10^{-35} $\,s,
i.e., the Compton time for $X$-particles); \
{(b)} long-living particles decay with $\tau_l > 1/M $.
    Remaind that in order to have $m_{X_l} < m_{X_s}$ one must take $A>0$
in Eq.~(7) of~\cite{GrPvAGN}.

     If $\tau_l$ is larger than the GU symmetry breaking time,
it can happen that some quantum number similar to a baryon charge can be
conserved, leading to some effective time
 $\tau_l^{\rm eff} > t_U \approx 4.3 \times 10^{17}$\,s
\  ($t_U $ is the age of the Universe).
    The small $ d \sim 10^{-15} - 10^{-12} $ part of long-living
$X$-particles with $\tau_l > t_U $ forms the dark matter.

For $ t_l^{\rm eff} \le 10^{27}$\,s one could have the observable flow of
UHECR from the decay in our Galaxy~\cite{BBV}.
But in this case one must get a strong anisotropy in the direction to
the center of the Galaxy~\cite{ABerK}.
However, Auger experiments do not show such an anisotropy, and one must
suppose $ t_l^{\rm eff} > 10^{27}$\,s.

    In our papers~\cite{GrPv}we considered observable entropy of the Universe
production due to decay of of $X$-particles.
To get the observable value of the entropy, it is sufficient to suppose the
interaction of the long-living component of $X$-particles with the baryon
charge (i.e. ordinary matter).
Then if we introduce the effective Hamiltonian of such interaction with
baryon matter as
     \begin{equation}
H^d =     \left(
\begin{array}{cc}
0  & 0   \\
0  & - i \gamma \\
\end{array}        \right)
\label{Hd}
\end{equation}
    one can write this parameter as being proportional to the concentration
of particles: $\gamma = \alpha\, n^{(s)}(t)$.
Then
     \begin{equation}
\alpha = \frac{ - 3 \ln d}{ 128 \pi (b^{(s)})^2} \,
\frac{M_{Pl}^2}{M^4} \,.
\label{ald}
\end{equation}
    For  $M=10^{14}$\,GeV  and  $d=10^{-14}$ one obtains
$\alpha \approx 10^{-40}$\,cm${}^2$.
    Thus such a mechanism of decay of the long-living component of
$X$-particles was important in the early Universe at time $ t_X$.
    In the modern era, due to smallness of the interaction cross-section
and concentration of $X$-particles, this mechanism of decay does not give
non-negligible contribution to the observable UHECR flow.

    The observed entropy in this scenario originates due to transformation of
$X$-particles into light particles: quarks, antiquarks and some particle
similar to $\Lambda^0$ in $K^0$-meson theory, having the same quantum
number as $X$.
Baryon charge is created close to the time $t_s $, which can be equal to
the Compton time of $X$-particles $t_C \sim 10^{-38} - 10^{-35} $\,s.

    Our scheme can also work for spinor particles.
Then it is possible to investigate some version of the {\it see-saw\/}
mechanism~\cite{GMRS} for Majorana neutrinos in GU theory,
so that heavy sterile neutrinos form the dark matter.

\vspace{11pt}
{\centering \section*{Acknowledgments}}

This work was supported by the Ministry of Science and Education
of Russia, grant RNP.2.1.1.6826.

\vspace{11pt}

\end{document}